\documentclass[twocolumn,showpacs,preprintnumbers,amsmath,amssymb,widetext]{revtex4}
\usepackage{graphicx}

\begin{document}
\title{Surface-acoustic-wave driven planar light-emitting device}
\author{Marco Cecchini}
\email{cecchini@sns.it}
\author{Giorgio De Simoni}
\author{Vincenzo Piazza}
\author{Fabio Beltram}
\affiliation{NEST-INFM and Scuola Normale Superiore, I-56126 Pisa,
Italy}
\author{H. E. Beere}
\author{D. A. Ritchie}
\affiliation{Cavendish Laboratory, University of Cambridge,
Cambridge CB3 0HE, United Kingdom}

%--------------------------------------------------------------
%--------------------------------------------------------------

\begin{abstract}
Electroluminescence emission controlled by means of surface
acoustic waves (SAWs) in planar light-emitting diodes (pLEDs) is
demonstrated. Interdigital transducers for SAW generation were
integrated onto pLEDs fabricated following the scheme which we
have recently developed \cite{cecchini-apl03}. Current-voltage,
light-voltage and photoluminescence characteristics are presented
at cryogenic temperatures. We argue that this scheme represents a
valuable building block for advanced optoelectronic architectures.
\end{abstract}

\pacs{72.50.+b,78.60.Fi,85.60.Jb,43.38.Rh}

\maketitle

Surface acoustic waves (SAWs) are attracting much interest in
semiconductor-device research owing to their interaction
properties with quasi two-dimensional systems (2DSs). Lattice
deformations induced by SAWs propagating on a piezoelectric
substrate (i.~e.~GaAs) are accompanied by potential waves which
interact with charge carriers confined in the heterostructure
leading to energy and momentum transfer. This interaction can drag
carriers along the SAW-propagation direction resulting in a net DC
current or voltage, the acoustoelectric effect
\cite{wixforth-surfsci94,wixforth-ssc92,campbell-ssc92}. Moreover,
this same interaction induces changes in SAW velocity and
amplitude that can be used to probe 2DSs transport properties
\cite{wixforth-prl86,willet-prl93,willet-prb92,simon-prb96}.

Several devices were proposed and realized exploiting the
acoustoelectric effect. Talyanskii \emph{et al.} proposed the
implementation of a novel current standard based on SAW-driven
transport through a quantum point contact
\cite{cunningham-prb00,cunningham-prb99,ebbecke-apl00,
talyanskii-prb97,talyanskii-jpcm96}, demonstrating very precise
quantization of the acousto-electric current down to
single-electron transfer per wave cycle. A particularly appealing
device proposal suggests the incorporation of a single-electron
SAW pump in a planar 2D electron/2D hole gas (n-p) junction to
realize a single-photon source\cite{talyanskii-pra00}.

In this Letter we demonstrate one of the main building blocks of
this latter single-photon source: we shall show transport and
optical emission in planar light-emitting n-p devices
(pLEDs)\cite{cecchini-apl03} controlled by means of SAWs. With the
pLED biased below threshold, the electric field associated with
SAW propagation in the GaAs facilitates the extraction and
transportation of electrons from the n-type region of the
structure into the 2D hole gas (2DHG) region, thereby generating
electroluminescence emission.

\begin{figure}
\includegraphics[width=8cm]{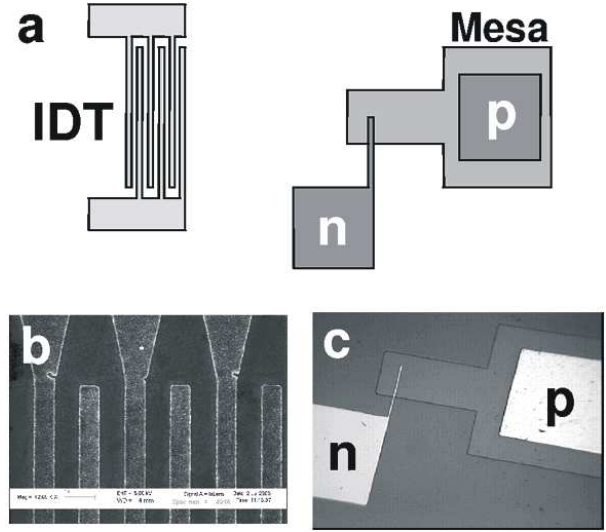}
\caption{(a) Schematic view of the SAW-driven light emitting
diode. (b) Scanning-electron-microscope image of part of the
transducer. (c) Optical photograph of the planar diode.}
\label{fig1}
\end{figure}

Devices were fabricated starting with a p-type modulation-doped
Al$_{0.3}$Ga$_{0.7}$As/GaAs heterostructure grown by molecular
beam epitaxy, containing a 2DHG within a 20-nm-wide GaAs quantum
well embedded 70\,nm below the surface. The measured hole density
and mobility after illumination at 1.5\,K were $2.0\times
10^{11}$\;cm$^{-2}$ and 35000\;cm$^{2}$/Vs, respectively. The
heterostructure was processed into mesas with an annealed p-type
Au/Zn/Au (5/50/150\,nm) Ohmic contact. The fabrication of the
n-type region of the junction started with the removal of the
Be-doped layer from part of the mesa by means of wet etching
(48\,s in H$_{3}$PO$_{4}$:H$_{2}$O$_{2}$:H$_{2}$O = 3:1:50) and
evaporation of a self-aligned Ni/AuGe/Ni/Au (5/107/10/100\,nm)
n-type contact. After annealing, donors introduced by the n-type
contact provide conduction electrons within the well. The
n-contact was shaped as a thin stripe placed perpendicular to the
SAW-propagation direction, 250\,$\mu$m away from the p-contact
(see Figs.~1a and 1c). SAWs propagating along the (0\={1}\={1})
crystal direction were generated by means of an interdigital
transducer (IDT) composed of 100 pairs of 200\,$\mu$m-long Al
fingers with 3\,$\mu$m periodicity ($\sim$1\,GHz resonance
frequency on GaAs). Transducers were fabricated at a distance of
800\,$\mu$m from the mesa by electron-beam lithography (see
Fig.~1b). The width of the n-type contact stripe (2\,$\mu$m) was
chosen to be of the same order of magnitude as the SAW wavelength
(3\,$\mu$m) in order to limit damping of the SAW electric field. A
schematic view of the device is shown in Fig.\,\ref{fig1}.

\begin{figure}
\includegraphics[width=8cm]{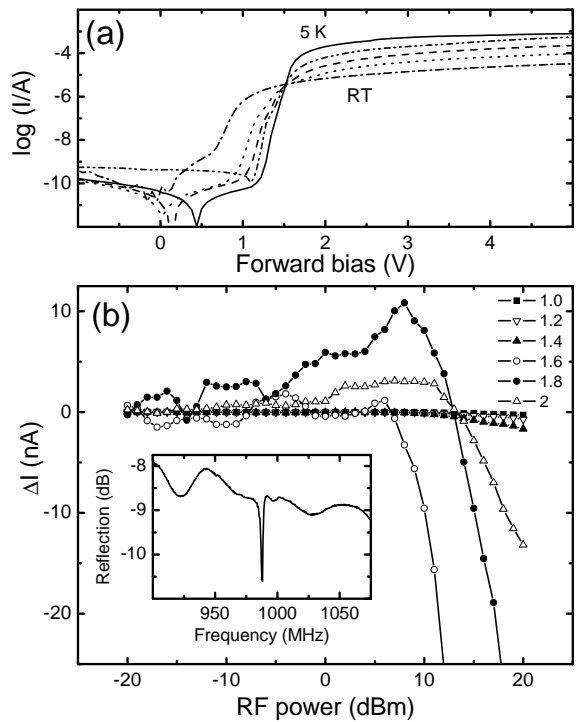}
\caption{(a) Current-voltage characteristics from room temperature
down to 5\,K. (b) Change of the current ($\Delta I$) flowing out
of the $n$-type contact with and without the presence of a SAW as
a function of the power of the RF signal applied ($P_{RF}$) to the
IDT at several forward biases around the device threshold
($\mathrm{T}=5\,\mathrm{K}$). Inset: Reflectivity of the
transducer as a function of the excitation frequency at
$\mathrm{T}=5\,\mathrm{K}$.} \label{fig2}
\end{figure}

Several pLEDs were fabricated with this protocol and preliminarily
electrically tested; all exhibited similar characteristics.
Fig.\,\ref{fig2}(a) shows the current-voltage (I-V) curves at
different temperatures of one representative device, on which all
the measurements reported in this Letter were taken. Rectifying
behavior is observed with a threshold voltage $\sim 1.5$\,V,
consistent with the value expected for a GaAs p-n junction.

The SAW resonance frequency was determined by measuring the power
reflected by the transducer as a function of excitation frequency.
The transducer frequency response (see inset of
Fig.\,\ref{fig2}(b)) displayed a dip at 987.5\;MHz with a full
width half maximum (FWHM) of 2.4\;MHz, at 5\;K, which is
consistent with the periodicity of the transducer.

The effect of SAWs on the device transport properties was studied
by measuring the change of the current ($\Delta I$) flowing out of
the n-type contact (with the p-type contact grounded) with and
without the presence of a SAW as a function of the power of the RF
signal applied ($P_{RF}$) to the IDT at several forward biases
around device threshold (see Fig.\,\ref{fig2}(b)). The RF signal
applied to the transducer was set at the measured resonance
frequency of 987.5\,MHz. $\Delta I$ was observed to increase (i.e.
more electrons injected into the 2DHG) for biases in the range of
1.6\,V to 2.0\,V. This regime was observed for values of $P_{RF}$
up to $\sim 10$\,dBm. At very high power levels ($P_{RF}
> \sim 10$\,dBm) a reduction in $\Delta I$ is observed probably
due to heat losses in the transducer and in the RF cable. These
measurements demonstrate that the acoustoelectric effect can
enhance the differential conductance of pLEDs near threshold.

\begin{figure}
\includegraphics[width=8cm]{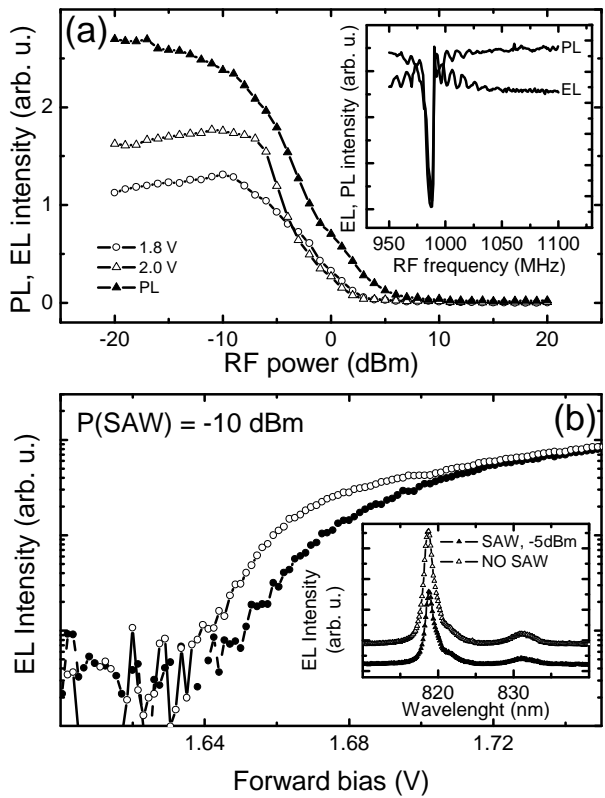}
\caption{(a) Photoluminescence (filled triangles) and
electroluminescence intensity as a function of $P_{RF}$
(987.5\,MHz) for two different forward voltages (1.8\,V empty
circles, 2\,V empty triangles). Inset: photoluminescence and
electroluminescence as a function of the RF-frequency (5\,dBm).
(b) Light-voltage characteristic of the pLED (filled circles) and
the same curve in presence a SAW (-10\,dBm, 987.5\,MHz). Inset:
electroluminescence spectra at a forward voltage of 2.0\,V with a
SAW (lower curve. $P_{RF} = -5$\,dBm) and without a SAW at
$\mathrm{T}=5\,\mathrm{K}$. The curves are vertically offset for
clarity.} \label{fig3}
\end{figure}

SAW control of the junction optical emission was demonstrated by
light-bias (L-V) measurements at low temperature (5\,K) (see
Fig.\,\ref{fig3}). Spectra as a function of bias were collected by
a cooled CCD after spectral filtering by a single-grating
monochromator. The electroluminescence (EL) spectra reported in
the inset of Fig.\,\ref{fig3}(b) show a main peak at 818.7\,nm
(FWHM 1.8\,nm) originating from radiative recombination within the
QW and a secondary peak at 831.2\,nm (FWHM 3.6\,nm) which is known
to originate from carbon impurities included in the
heterostructure material during growth\,\footnote{B. Hamilton, in
\emph{Properties of Gallium Arsenide}, edited by M. R. Brozel and
G. E. Stillman (INSPEC, London, England, 1996).}. Furthermore, the
SAW can be seen to not change significantly the shape of the
spectra.

Figure \ref{fig3}(a) shows the effect of SAWs on the optical
properties of the pLEDs. Light intensity, calculated by
integration over the main EL peak from 810\,nm to 826\,nm, is
reported as a function of $P_{RF}$ (at transducer resonance
frequency, 987.5\,MHz) for two different values of forward bias.
Two different regimes can be observed: an increase in EL intensity
up to $P_{RF}\sim -5$ dBm followed by an abrupt suppression of the
emission. The latter regime can be linked to the spatial
separation between electrons and holes induced by the SAW electric
field \cite{wixforth-prl97,alsina-prb01}, which is also observed
in the PL measurements\,\footnote{PL spectra were obtained by
excitation of a region of the mesa with a red-light laser source
(653\,nm).} (filled triangles in Fig\,\ref{fig3}(a)). Remarkably,
the increase in intensity observed for $P_{RF}< -5$ dBm is unique
to the EL measurements as it was not observed in the PL data. The
resonant nature of this effect is shown in the inset of
Fig.\,\ref{fig3}(a) where the light intensity is plotted as a
function of the frequency of the signal applied to the IDT
(RF-frequency) for $P_{RF}=5$\,dBm. A clear dip in both EL and PL
data is present at the observed transducer resonant frequency.
These observations and the measured increase of forward current in
presence of SAWs demonstrate SAW-driven injection of electrons
into the 2DHG leading to increased EL intensity.

This is further highlighted by the comparison between L-V data
collected with and without SAWs shown in Fig\,\ref{fig3}(b), where
the EL threshold is reduced from 1.65\,V to 1.64\,V in the
presence of SAWs.

In conclusion we demonstrated the possibility of controlling the
electroluminescence emission from a planar diode by means of
surface acoustic waves. Planar-junction devices with interdigital
transducers were fabricated and characterized by transport and
optical measurements at cryogenic temperatures. We have
demonstrated that SAWs can induce  transport and light emission in
these devices. SAW-driven light-emitting diodes are one of the
fundamental building blocks of the single-photon source proposed
in Ref.~\cite{talyanskii-pra00}. This work was supported in part
by the European Commission through the FET Project SAWPHOTON and
by MIUR within FISR ``Nanodispositivi ottici a pochi fotoni''. We
thank M. Lazzarino for technical help in device fabrication.

%------ References --------------------

\bibliography{sawled}

%------ Figures beginning -------------

\end{document}